\documentclass[RNAAS]{aastex631}

\usepackage[edges]{forest}
\usepackage{tikz}
\usetikzlibrary{fit,positioning}
\usepackage{bayesnet}  

\usepackage{verbatim, amsmath, amsfonts, amssymb,amssymb}
\usepackage{multirow}
\usepackage[utf8]{inputenc}
\usepackage{empheq}
\usepackage[skins,theorems]{tcolorbox}
\usepackage{latexsym}
\usepackage{bigints}
\usepackage{natbib}
\usepackage{color}
\usepackage{lipsum}
\RequirePackage[fixed]{fontawesome5}
\usepackage{tikz}
\usepackage{hyperref}

\newcommand{\github}[1]{%
   \href{#1}{\faGithub}%
}
\shorttitle{Bayesian Globular Cluster Counts}
\shortauthors{de Souza et al.}
\graphicspath{{./}{figures/}}

\begin{document}
\usetikzlibrary{positioning}

\title{Modeling Globular Cluster Counts with Bayesian Latent Models}

\correspondingauthor{Rafael S. de Souza}
\email{rd23aag@herts.ac.uk, drsouza@ad.unc.edu}

\author[0000-0001-7207-4584]{Rafael S. de Souza}
\affiliation{Centre for Astrophysics Research, University of Hertfordshire, College Lane, Hatfield, AL10~9AB, UK}
\affiliation{Instituto de Física, Universidade Federal do Rio Grande do Sul, Porto Alegre, RS 90040-060, Brazil}
\affiliation{Department of Physics \& Astronomy, University of North Carolina at Chapel Hill, NC 27599-3255, USA}
\author[0000-0000-0000-0002]{Ana L. Chies-Santos} 
\affiliation{Instituto de Física, Universidade Federal do Rio Grande do Sul, Porto Alegre, RS 90040-060, Brazil}


\begin{abstract}
We present a Bayesian latent model to describe the scaling relation between globular cluster populations and their host galaxies, updating the framework proposed in \citet{desouza2015}. GC counts are drawn from a negative-binomial (NB) process linked to host stellar mass, augmented with a newly introduced Gaussian observation layer that enables efficient propagation of measurement errors. The revised formulation preserves the underlying NB process while improving computational tractability. The code snippets, implemented in \texttt{Nimble} and \texttt{PyMC} are released under the MIT license at \href{https://github.com/COINtoolbox/Generalized-Linear-Models-Tutorial/blob/master/Count/readme.md}{this repository \faGithub}.
\end{abstract}



\definecolor{anorange}{RGB}{230,120,20} 
\newcommand{\acs}[1]{\textcolor{anorange}{#1}}
\section{Introduction} 

Globular clusters (GCs) serve as valuable tracers of the formation histories and global properties of their host galaxies. Empirical correlations have long been observed between the number of GCs ($N_{\rm GC}$) and various  host-galaxy properties, including stellar mass, halo mass, and central velocity dispersion \citep[e.g.][]{blakeslee97,Harris2013,hudson14,val21,zar22,saif22, Canossa2024, dornan25} %
For systems beyond the Local Group, $N_{\rm GC}$ are typically more accessible than their total masses, making the scaling relation between the number of clusters  and galaxy mass of particular practical and physical interest.

These relations are often fitted using linear regressions intended for continuous responses, despite the inherently discrete nature of count data. Count models—members of the generalized linear model  family—provide a principled alternative by linking a linear predictor to a discrete response, ensuring integer-valued outcomes and more realistic uncertainty quantification, especially in the low-count regime. Yet such models remain underutilized in astrophysical research, even though many observables—photon counts, supernova events, and planetary detections among them—are intrinsically discrete. 

In \citet{desouza2015}, we introduced a Bayesian negative-binomial (NB) framework—an exponential family distribution under the umbrella of generalized linear models \citep[see e.g.,][]{souza2015,Elliott2015}—to describe the dispersion of GC populations across a broad range of galaxy properties. This work, together with \citet{hilbe2017} and \citet{Hattab2019}, in which we implemented a hurdle model to characterize the escape fraction of ionizing radiation during the Epoch of Reionization, has since motivated a series of studies that expanded upon our original approach \citep{Berek2023,Berek2024}. The present note revisits that framework, refining the treatment of discrete measurement errors in GC counts.
The revised model retains the hierarchical latent-variable treatment for photometric uncertainties but adopts a Gaussian observation layer over the NB process, providing a computationally efficient approximation that remains faithful across most of the observed range. 
\clearpage
\section{Dataset}
\label{sec:catalog}

To showcase the new model, we use a public catalog of globular cluster systems \citep{Harris2013,harris14}\footnote{\url{http://www.physics.mcmaster.ca/~harris/GCS_table.txt}} comprised by 422 galaxies covering the full Hubble sequence, 247 ellipticals, 94 lenticulars, 55 spirals, and 26 irregulars, for which both $N_{\mathrm{GC}}$ and auxiliary photometric quantities are available. 
For the present analysis, we adopt the stellar mass as the primary predictor, estimated from the $V$-band absolute magnitude and $(B\!-\!V)$ color using the mass-to-light ratio prescription of \citet{Bell2003}, and only fitted the elliptical galaxies. 
The resulting masses span approximately $7.5 \lesssim \log_{10}(M_\star/M_\odot) \lesssim 12$.

\section{Model}

The updated framework models the observed globular-cluster counts and stellar masses, for galaxies \(i = 1, \dots, N\), as noisy measurements of latent true quantities within a hierarchical Bayesian structure.

The updated framework models the observed globular cluster counts and stellar masses as noisy measurements of latent true quantities within a hierarchical Bayesian structure. 

At the core, the intrinsic counts are governed by a discrete negative–binomial (NB) process,
\begin{align}
N_{\mathrm{GC},i} &\sim \mathrm{NB}(p_i, k), &
p_i &= \frac{k}{k+\mu_i}, \notag\\
\mu_i &= \exp(\eta_i), &
\eta_i &= \beta_0 + \beta_1\, x_i, \qquad x_i \equiv \log_{10}\!\big(M_{\star,i}^*/M_{\odot}\big), \notag\\
x_i^{\mathrm{obs}} &\sim \mathcal{N}\!\big(x_i,\, e_{x,i}^2\big), &
N_{\mathrm{GC},i}^{\mathrm{obs}} &\sim \mathcal{N}\!\big(N_{\mathrm{GC},i},\, e_{N_{\mathrm{GC}},i}^2\big).
\end{align}
Here, $x_i=\log_{10}(M_{\star,i}^*/M_{\odot})$ is the latent (true) stellar mass in solar units; $e_{x,i}$ and $e_{N_{\mathrm{GC}},i}$ are the reported observational uncertainties. Although Gaussian terms describe measurement noise (appropriate for continuous quantities), the NB likelihood preserves the discreteness of $N_{\mathrm{GC}}$. The dispersion parameter $k$ controls departures from the Poisson limit.
This formulation allows measurement errors in both axes to be explicitly modeled, preserving the discreteness of the intrinsic GC counts while accommodating observational scatter.

\section{Analysis and Discussion}
\begin{figure}[ht!]
\plotone{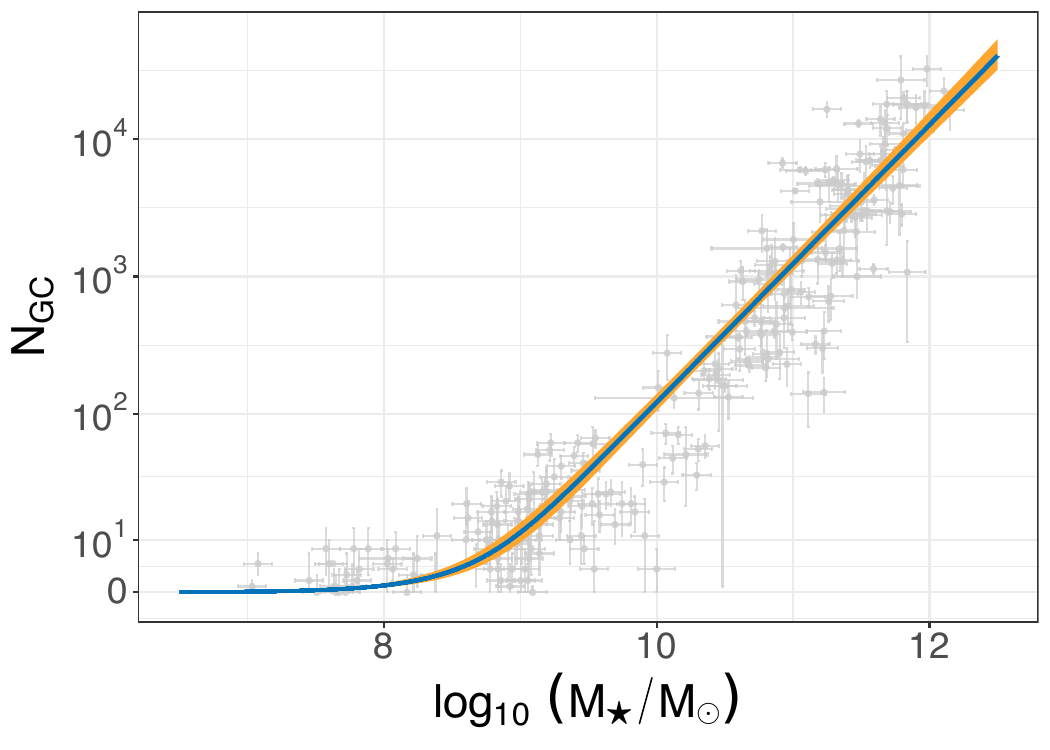}
\caption{Negative–binomial regression with a latent errors-in-variables treatment for the relation between $N_{\rm GC}$ and $\log_{10}(M_{\star}/M_{\odot})$.
Data points are elliptical galaxies from \citet{Harris2013}. The solid blue curve shows the posterior median, and the orange band the 95\% credible interval. Points include 1$\sigma$ uncertainties. }
\label{fig1}
\end{figure}

We fit the hierarchical negative–binomial model using
$x=\log_{10}(M_{\star}/M_{\odot})$ as the predictor derived from $M_V$ and color (Section~\ref{sec:catalog}). 
The fitted relation and uncertainties are shown in \autoref{fig1}. 
The NB likelihood accommodates strong extra-Poisson variation while preserving the discreteness of $N_{\rm GC}$, so zeros and small counts are handled naturally. 
Across the sampled mass range the trend is positive and approximately log-linear; the 95\% credible band encloses the full dynamic range, including systems with $N_{\rm GC}=0$. 
Posterior checks (not shown) reveal no large-scale residual structure, and the inferred dispersion parameter $k$ confirms substantial overdispersion relative to a Poisson model.

\autoref{fig1} shows the empirical relation: the model preserves the lower \emph{bound} of GC counts and properly propagates observational uncertainties through the latent, discrete NB layer. 
The $y$ axis uses the base-10 pseudo-log transform with $\sigma=5$, 
$y'=\operatorname{asinh}\dfrac{(y/5)}{\ln 10}$, which is linear near zero and logarithmic at large $y$, providing a continuous scale that accommodates both small and zero counts without distortion.

\begin{acknowledgments}
RSS acknowledges support from the Conselho Nacional de Desenvolvimento Científico e Tecnológico (CNPq, Brazil), process nos. 446508/2024-1 and 315026/2025-1.
ACS acknowledges support from FAPERGS (grants 23/2551-0001832-2 and 24/2551-0001548-5), CNPq (grants 314301/2021-6, 312940/2025-4, 445231/2024-6, and 404233/2024-4), and CAPES (grant 88887.004427/2024-00).

\end{acknowledgments}

%

\vspace{5mm}
%



\software{Nimble \citep{deValpine2017}, PyMC \citep{Salvatier2015}}.

\bibliography{ref}{}
\bibliographystyle{aasjournal}



\end{document}